\documentclass[a4paper]{jpconf}
\usepackage{graphicx}
\begin{document}
\title{Recent {\it MOST}\footnote{Based on data from the MOST satellite, a Canadian Space Agency mission, jointly operated by Dynacon Inc., the University of Toronto Institute of Aerospace Studies and the University of British Columbia with the assistance of the University of Vienna.} space photometry}

\author{Gordon A.H. Walker}

\address{Dept. Physics \& Astronomy, UBC, 
 Vancouver, BC V6T 1Z1, Canada}

\ead{gordonwa@uvic.ca}

\begin{abstract}
The {\it Microvariability and Oscillations of STars ({\it MOST})} photometric satellite has already undertaken more than 64 primary campaigns which include some clusters and has obtained observations of $>$850 secondary stars of which $\sim$180 are variable. More than half of the variables pulsate, with the majority being of B-type. Since 2006 January, MOST has operated with only a single CCD for both guiding and science. The resulting increase in read-out cadence has improved precision for the brightest stars. The 2007 light curve for Procyon confirms the lack of predicted p-modes with photometric amplitudes exceeding 8 ppm as we found in 2004 and 2005. p-modes have been detected in other solar-type stars as well as pre-main sequence objects, roAp and $\delta$ Scuti variables. g-modes have been detected in a range of slowly pulsating B stars, Be stars and $\beta$ Cephei variables. Differential rotation has been defined for several spotted solar-type stars and limits set to the albedo of certain transiting planets and the presence of other perturbing planets. The mission is expected to continue as long as the experiment operates. 
\end{abstract}

\section{{\it MOST}}
The {\it MOST} photometric satellite, originally proposed by Slavek
Rucinski, was launched on 30 June 2003 into a 101.4 min sun-synchronous
orbit (820 km) at a total cost of C\$10M. The experiment, more fully described by Walker et al. (2003), consists of a
15/17.3 cm Rumak-Maksutov telescope feeding two CCDs, one for tracking, the other
for science, through a single broadband filter
(350 -- 700 nm). In the Fabry mode, an image of the telescope entrance pupil covering some 1500 pixels is projected onto the science CCD. An area of the
science CCD is also devoted to in-focus photometry of fainter stars.

The primary goal was detection of solar-type p-modes with periods of minutes at micro-magnitude precision. The engineers achieved a dramatic reduction in
tracking jitter early in 2004 to $\sim$1 arcsec (essentially diffraction limited performance) leading to a significant improvement in Direct imaging photometric precision and recognition that precise photometry was also possible from the guide star counts returned by the tracking CCD. This opened up a much larger number, and broader range, of variables for observation. Exposures of up to 1 minute were possible in both the Fabry and Direct fields with $\sim$1 sec for guide stars.

Early in 2006, the tracking CCD system failed, likely from a particle hit. Since then, both science and tracking have been carried out with the Science CCD which limits Fabry and Direct photometric exposures to between 0.1 and 3 sec.
On the other hand, there has been a significant improvement in precision for the brightest stars because of a higher read-out cadence coupled with an absence of intermittent x-talk from the tracker electronics. The cost of this improvement is the loss of candidates in the 3.5 - 6  mag range because they are too faint for short exposures in Fabry mode but too bright for the in-focus Direct mode.

The presence of parasitic light, mostly Earth shine, at certain orbital phases has posed the greatest challenge in the data analysis. The amount of the background depends on the stellar coordinates, spacecraft roll and season of the year.
Neutralising the impact of the periodic parasitic light involved a considerable effort at both the UBC Physics \& Astronomy Department and the Institute f\"ur Astronomy, Wien for both the Direct and Fabry photometry (see Rowe et al. 2006, and Reegen et al. 2006). But, with this problem finally solved, the pace of paper writing has greatly increased and a list can be found at http://www.astro.ubc.ca/MOST/science.html. I can only touch on a very few highlights in this brief review.

\section{some observing statistics}

As of June 2007, {\it MOST} had been in operation for four years, with scientific observations having been made  for 3.5 years. There were 64 campaigns which also included some clusters and data were recorded for $>$850 secondary stars, of which $\sim$180 have proved to be variable. Figure 1 is a histogram of the \% stars as a function of spectral type. 

\begin{figure}[!ht]
\begin{center}
\includegraphics[width=20pc]{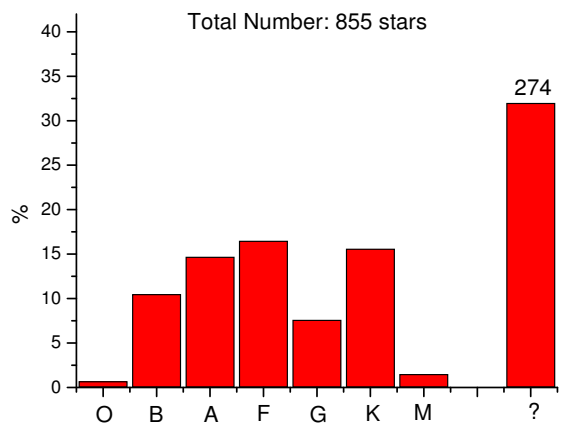}\hspace{2pc}
\begin{minipage}[b]{14pc}\caption{\label{label}Histogram of the target and secondary stars observed by {\it MOST} up to June 2007. (Figure prepared by Rainer Kuschnig).}
\end{minipage}
\end{center}
\end{figure}

Of the 180 variables: 100 pulsate, 17 are rotationally modulated, 15 are eclipsing binaries and 53 show long-term periods or trends.

Among the 100 pulsators: 34 are B stars ($\beta$ Ceph, SPB, SPBe and SPBsg), 26 $\delta$ Scuti, 18 $\gamma$ Dor, and 22  K-G giants.

Considerable energy will go into the archiving of these and up-coming data which will be an important resource for years to come.

\section{Procyon and $\alpha$ CenA}

The {\it MOST} specifications were based on the key search for the $\sim$15 min p-mode oscillations predicted for the F5 IV star Procyon - one of our earliest targets. Matthews et al. (2004) reported that no such p-modes were detected in the 32 day 2004 light curve. This null result attracted a lot of attention and some were skeptical (eg Bedding et al. 2005, R\'egulo and Roca Cort\'es 2005, and Bruntt et al. 2005).  But, subsequent photometry of increasingly high precision with {\it MOST} has upheld this result and better defined the limits.

\vspace{8 mm}
\begin{figure}[!ht]
\begin{center}
\includegraphics[width=3in]{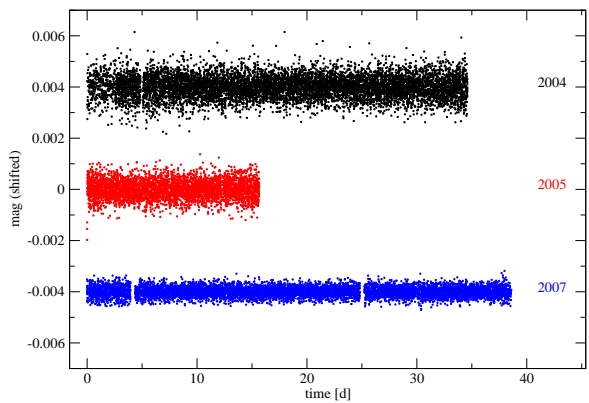}\hspace{2pc}
\begin{minipage}[b]{14pc}\caption{\label{label}Light curves from the \emph{MOST} 2004, 2005, \& 2007 Procyon runs. Only every tenth datapoint is plotted. (Figure courtesy Daniel Huber).}
\end{minipage}
\end{center}
\end{figure}

\vspace{4 mm}
\begin{figure}[h]
\includegraphics[width=6.0in, angle=0]{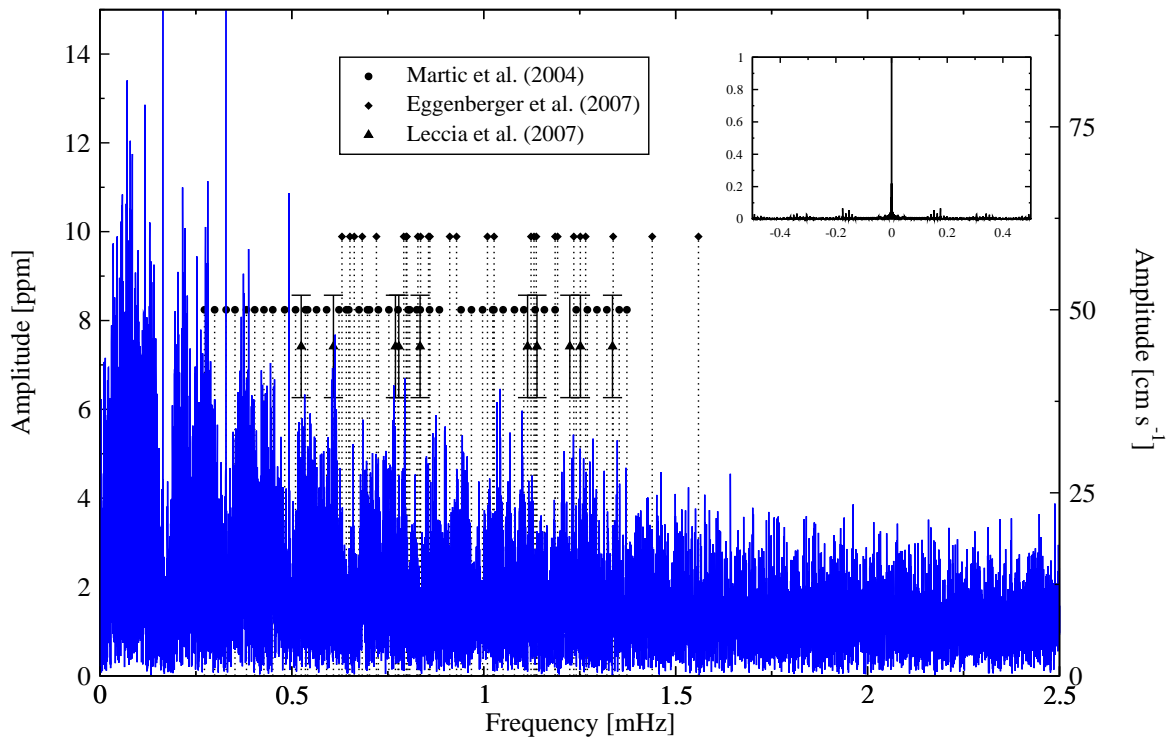}
\vspace{5 mm}
\caption{Procyon \emph{MOST} amplitude spectrum from 2007. Symbols indicate published spectroscopic frequencies and amplitudes. Velocity amplitude was
converted to photometric amplitude with the relation from Kjeldsen \& Bedding (1995). (Figure courtesy Daniel Huber).} 
\label{label}
\end{figure}

The light curves from 2004, 2005 and 2007 are shown in Figure 2. The 2005 run, although only half as long benefited from the improved pointing and had a $\sigma\sim$350 ppm, from individual 1 sec exposures taken every 30 sec. This year's 38 day run had the added benefit of accumulation of images aboard the space craft with individual exposures of 0.8 sec and 21 stacked exposures read out every 16.8 sec. The result being a photon detection gain of $\sim$30 over earlier years and a reduction to $\sigma\sim$200 ppm. Despite this improvement the predicted p-mode spectrum is not seen in the amplitude spectrum from 2007 in Figure 3 where the noise level is $\sim$2 ppm and the amplitude predicted from spectroscopic observations should be $\sim$8 ppm based on the conversion of velocity amplitude to photometric amplitude given by Kjeldsen \& Bedding (1995). 

None the less, there are several peaks with S/N$>$4 within the 1 to 1.5 mHz range where p-modes are expected. But these peaks do not show the regular frequency spacing expected for such modes.  We confirmed that these peaks are not of instrumental origin by comparison with {\it MOST} observations of Arcturus
also from 2007 which shows no such features $<$ 1.5\,mHz. An analysis of the results and interpretation using 3D numerical  
simulations of Procyon's convective envelope (Robinson et al. 2005)  
is in progress.

Evidence for p-modes has been found in {\it MOST} observations of 
the G2 V star $\alpha$ CenA. $\alpha$ CenB (K1 V $\Delta$m=1.4 at 21 asec) was included in the field stop. The average noise level between 1.5 - 3.8 mHz
was $\sim$2.5-3.5 ppm. Several of the 19 most significant amplitude peaks coincide in frequency with those found from spectroscopy. The spectroscopic peaks have already been successfully modeled by David Guenther.

\section{Giants}

{\it MOST} has observed specific red and yellow giant target stars as well as detecting a number in the Direct field for the old open cluster M67 in 2007. This is an on-going program and the potential of the red giant light curves is well illustrated by $\epsilon$ Oph. 

{\it MOST} observed the G9.5 giant $\epsilon$ Oph for 28 days in 2005 and both Barban et al. (2007) and Kallinger et al. (2007) have analysed the light curve. 
Kallinger et al. (2007) find clear evidence of non-radial p-modes in both radial velocity and luminosity. Their best fitting model to 18 of the 21 most significant photometric frequencies lies within $\pm1\sigma$ of $\epsilon$ Oph known position in the HR-diagram and its interferometrically determined radius. The comparison is given in Table 1.

\begin{table}[h!]
\begin{center}
\caption{Fundamental stellar parameters for $\epsilon$ Oph found  
in the literature and for the best model fit to the {\it MOST} light curve. (Table courtesy Thomas Kallinger).
\label{tab2}}

\begin{tabular}{lcc}
\hline
\hline
\noalign{\smallskip}
& Literature & best fitting \\
&&model\\
\noalign{\smallskip}
\hline
\noalign{\smallskip}
Effective temperature  ...  [K] &4877$\pm$100$^1$ &4892\\
Luminosity  ...  [L$_\odot$] &59$\pm$5$^1$ &60.13\\
Radius  ...  [R$_\odot$] &10.4$\pm$0.45$^2$&10.82\\
Mass  ...  [M$_\odot$] & &2.02\\
log g  ...  [g cm s$^{-2}$] &2.48$\pm$0.36$^3$&2.674\\
Age  ...  [Gyr] & &0.770\\
Mixing length parameter  ...  [Hp] & &1.74\\
Metallicity  ...  [Z] &0.01$^4$ &0.01\\
\noalign{\smallskip}
\hline
\noalign{\smallskip}
\multicolumn{3}{l}{$^1$De Ridder et al. (2006)}\\
\multicolumn{3}{l}{$^2$Richichi et al. (2005) and Hipparcos}\\
\multicolumn{3}{l}{$^3$Allende \& Lambert (1999)}\\
\multicolumn{3}{l}{$^4$[Fe/H] = -0.25 from Cayrel de Strobel et al.  
(2001)}\\
\end{tabular}
\end{center}
\end{table}

\section{Pulsating pre-main sequence stars}

{\it MOST} has targeted a number of pulsating, pre-main sequence (PMS) stars and discovered others in the Direct field for such associations as NGC 2264. Despite large, irregular light variations caused by transiting dust clouds, clear pulsational signals have been detected from many of them. This is illustrated by the light curves in Figure 4 for the A8 Ve PMS star HD 142666 (V 1026 Sco) as observed by {\it MOST} in 2006 and 2007. Figure 5 highlights the most significant frequencies detected in each year dominated by peaks at 21.253 and 22.009 c d$^{-1}$ (1.13 and 1.09 h) with seven of the most significant peaks being in common between the two years. Initial modeling confirms that the detected frequencies can only be explained with pre-MS [as opposed to post-] pulsation models thereby providing direct evidence for the PMS evolutionary stage. This work is in progress. 

\vspace{10 mm}
\begin{figure}[h]
\begin{center}
\includegraphics[width=5.0in, angle=0]{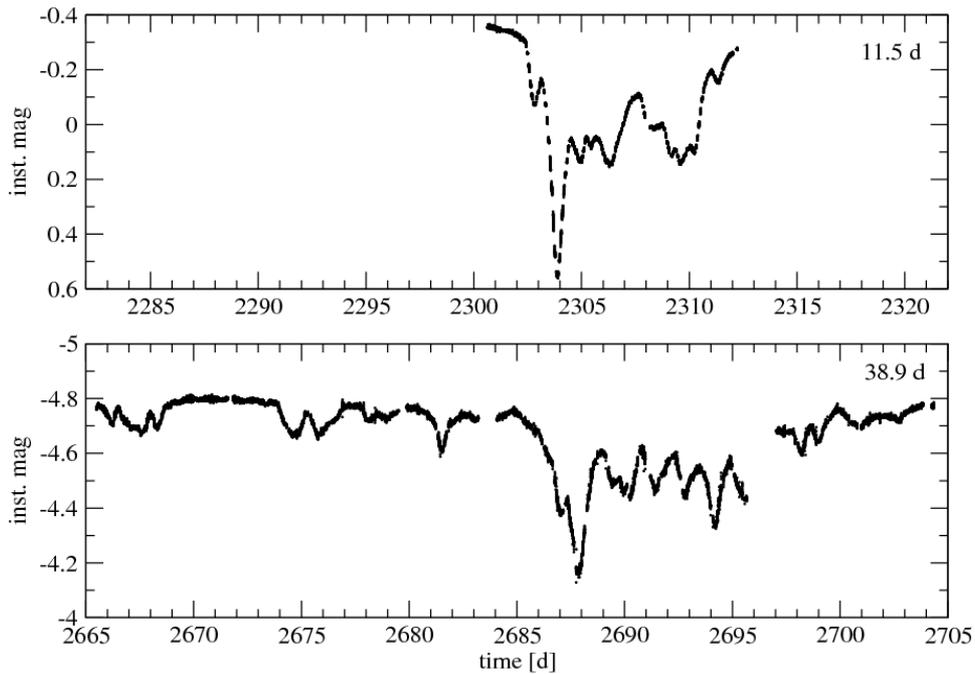}
\vspace{5 mm}
\caption{The 2006 and 2007 {\it MOST} light curves of the pre-main sequence star HD 142666. The large variations are caused by transiting dust clumps in the dense surrounding disc. (Figure courtesy of Konstanze Zwintz).} 
\label{label}
\end{center}
\end{figure}

\vspace{3 mm}
\begin{figure}[h]
\begin{center}
\includegraphics[width=4.5in, angle=0]{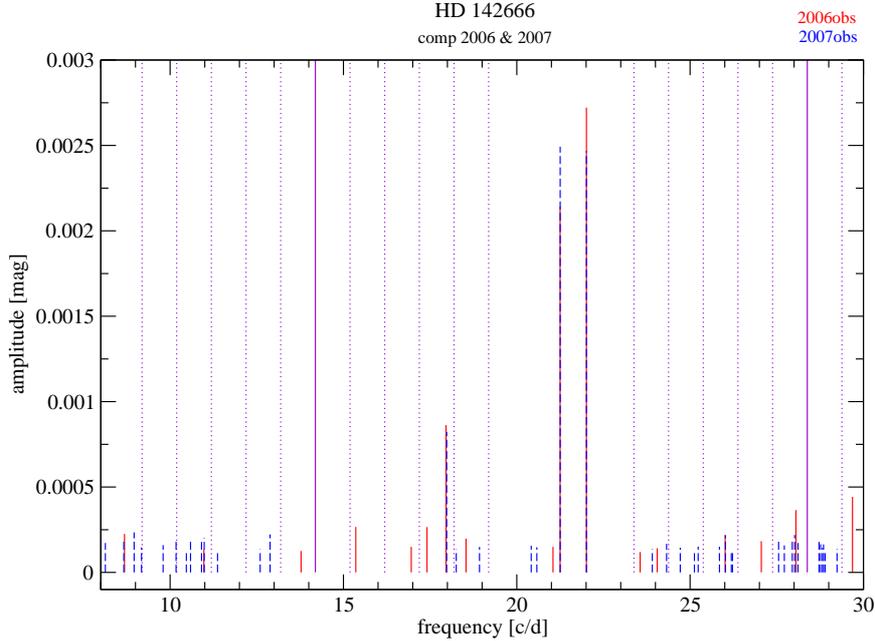}
\vspace{5 mm}
\caption{ Significant frequencies for HD 142666 from 2006 are shown by solid red - 2007 by blue dashed lines. The violet lines are the {\it MOST} orbital period, its harmonics and 1 d$^{-1}$ 
aliases. (Figure courtesy of Konstanze Zwintz).}
\label{label}
\end{center}
\end{figure}

\section{roAp stars}

{\it MOST} has observed several roAp stars.

The rapidly oscillating Ap (roAp) star HR 1217 (HD 24712)
is a prime candidate for testing the interaction between
a stellar magnetic field and nonradial acoustic (p-mode)
pulsations.  HR 1217 exhibits a rich eigenspectrum of
frequencies which obey the expected asymptotic relation
for p-modes of low degree and high overtone, and show
fine splitting due to rotational modulation. A global
multisite photometric campaign (Whole Earth Telescope) on
this star revealed an anomalous frequency consistent with a magnetically perturbed mode, as
predicted theoretically by Cunha (2001) and evidence for a small mode spacing (Kurtz et
al. 2002 and 2005).

{\it MOST} monitored HR 1217 for almost 30 days
in late 2004.  Analysis confirms the
results of Kurtz et al. (2005) with high accuracy
and identifies additional pulsation frequencies.
This leads to spacings consistent with the range of expected
small spacings ($\sim 3\mu$Hz) for a main sequence A-F star.
A grid of stellar evolution and pulsation models to estimate the  
effect of dipolar magnetic fields of various strengths on the  
observed frequencies and mode spacings are being tested by C. Cameron  
as part of his Ph.D. thesis at UBC.

\section{B stars}

One unanticipated success of {\it MOST} has been the serendipitous discovery of a large number of slowly pulsating B (SPB) stars among guide stars and in the direct imaging field. The high precision and continuity of the light curves has made them highly amenable to modeling. Waelkens (1991) first coined the term `slowly pulsating B star' (SPB) for a group of B8 to B3 stars of 3 to 7 $M_\odot$ with multiperiodic light and color variations. Their periods of several hours were much too long for radial pulsations. The variations can be explained by $g$-mode pulsations of low spherical harmonic degree ($\ell$) and high radial order ($n$) excited by the $\kappa$-mechanism in the Fe-opacity bump. Originally thought to be slow rotators, Aerts et al. (1999) found several rapidly rotating SPBs with more complex line profile variations.

Figure 6 shows our first published example (see Walker et al. 2005b) of a {\it MOST} SPB light curve for HD 163868 together with a successful model by Hideyuki Saio in Figure 7. The star rotates rapidly ($v_{\rm e} = 305$ km $s^{-1}$) corresponding to a rotational frequency of 0.016 mHz. Prograde g-modes with $m=-1$ then have observed frequencies which are the sum of the rotational frequency and the frequency in the corotating frame, while those with $m=-2$ have double the frequency. These two well separated groups of frequencies are a characteristic of the SPB stars.

Table 2 lists the variable B stars already `solved' by {\it MOST} and modeling invoking the Fe opacity bump and those in process. The list includes several $\beta$ Cephei cadidates. Lef\`evre et al. (2005) detected a persistant 9.8 h periodicity in the WN8 star WR 123 which might be related to g-modes (see Noels in these proceedings).

\vspace{15 mm}
\begin{table}[h!]
\caption{\label{label}B star variables `solved' by {\it MOST} and the Fe opacity bump.}
\begin{center}
\begin{tabular}{llll}
\br
~ star &spectrum &variable &status\\
\mr
(WR 123   & WN8 & 9.8 h period?)&Lef\`evre et al. 2005\\
$\zeta$ Oph   &  Oe & $\beta$ Cephei&Walker et al. 2005a\\
Spica        &   B1 IV     &  $\beta$ Cephei & in process\\
HD 127756    &  B2 Vne  &   SPBe  &   in process\\
$\delta$ Ceti     &   B2 IV   &    $\beta$ Cephei& Aerts et al. 2006a\\
HD 163899   &   B2 Ib/II  &  SPBsg&Saio et al. 2006\\
HD 163868  &    B2.5 Ve  &   SPBe& Walker et al. 2005b\\
HD 217543   &   B3 Vpe    &  SPBe   &  in process\\
HD 163830   &  B5 II/III  &  SPB&  Aerts et al. 2006b \\
$\beta$ CMi       &       B8 Ve   &     SPBe&Saio et al. 2007\\
\br
\end{tabular}
\end{center}
\end{table}
\vspace{10 mm}

\begin{figure}[h!]
\begin{center}
\includegraphics[width=4.5in, angle=0]{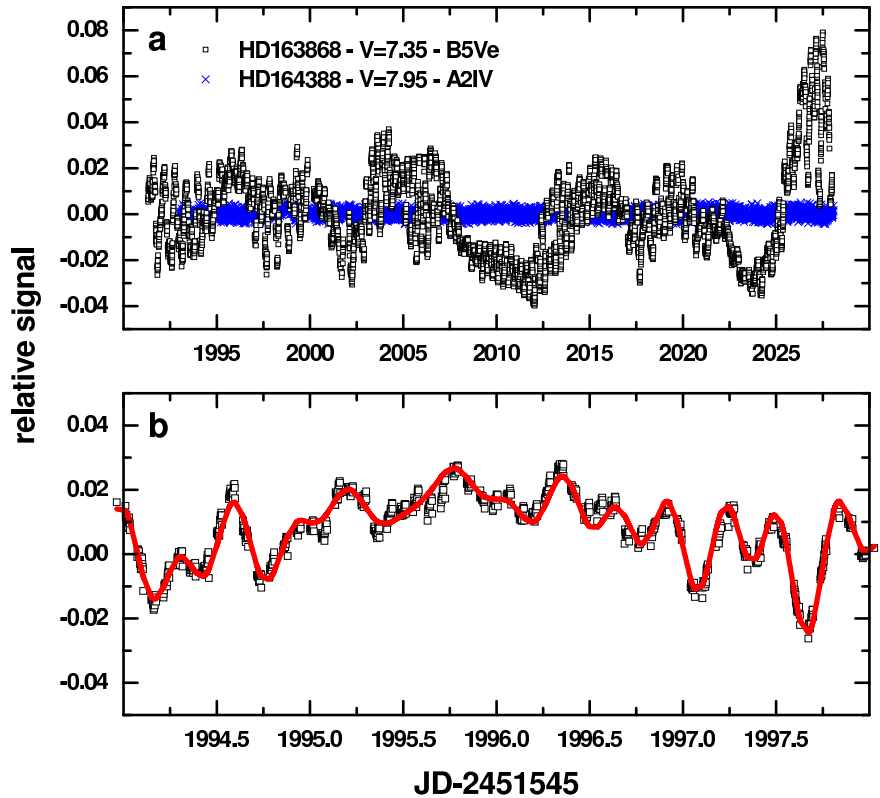}
\vspace{3 mm}
\caption{{\it upper:} The 2005 {\it MOST} 37 day light curve of the guide star HD 163868 and the  simultaneously observed guide star HD 164388; {\it lower:} a 5 day portion of the HD 163868 light curve with the data binned every 2 min. The red line is the fit of the sixty detected frequencies to the data. (Figure from Walker et al. 2005b). } 
\label{label}
\end{center}
\end{figure}

\begin{figure}[h!]
\begin{center}
\includegraphics[width=5.0in, angle=0]{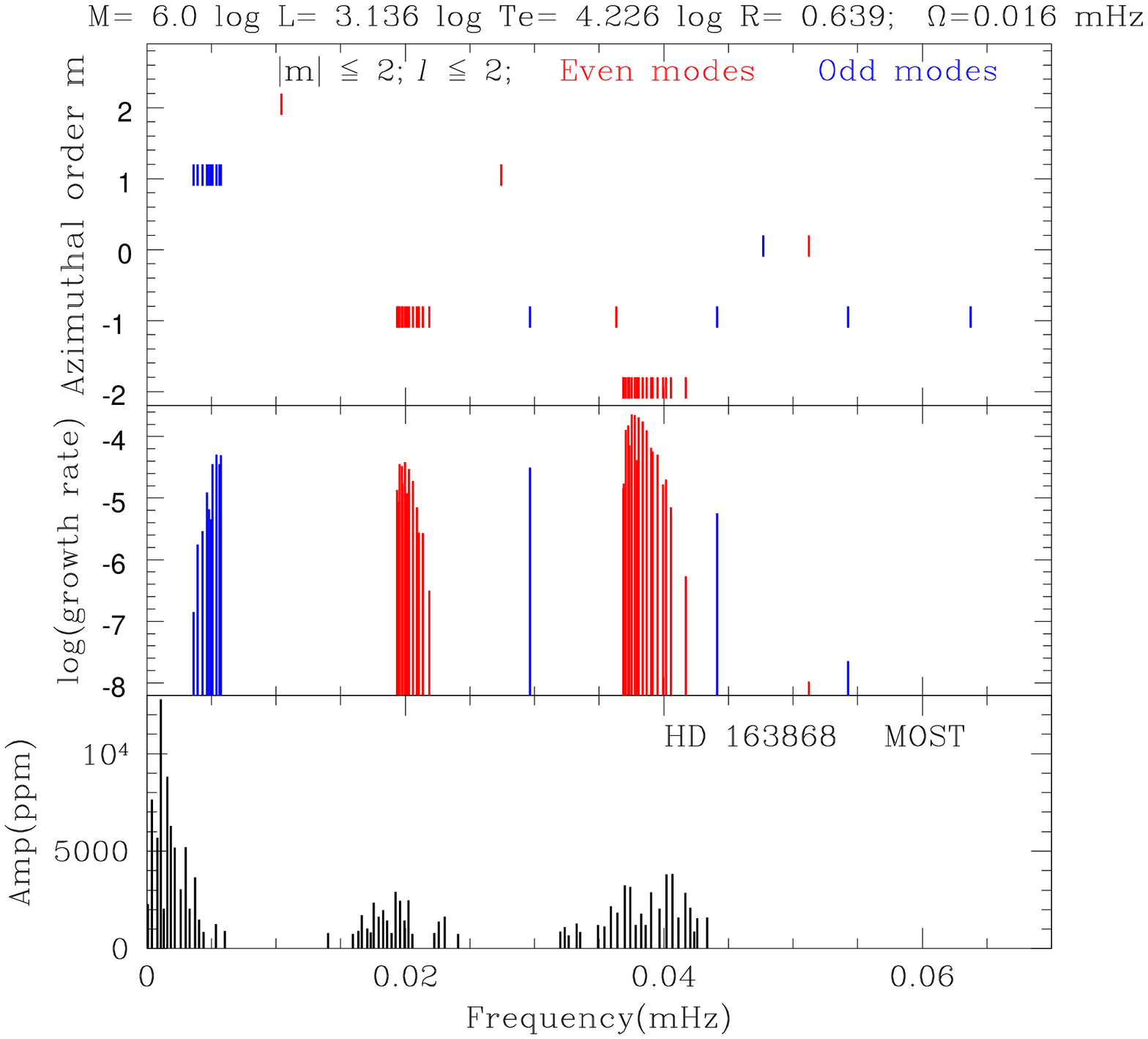}
\caption{Top and middle panels: frequencies and growth rates of excited
pulsations in a $6M_\odot$ main-sequence model with a rotational frequency
of 0.016 mHz ($v_{\rm e} = 305$ km s$^{-1}$).  Only low latitudinal
degree ($\ell \le 2$) and low azimuthal order ($|m|\le 2$) modes are
shown.  Red and blue lines refer to even and odd modes.  Odd modes at $\sim 0.005$ mHz are thought to be high order r-modes, while most of the others are high order g-modes. Bottom panel: HD 163868 amplitude spectrum from {\it MOST} (Figure from Walker et al. 2005b).} 
\label{label}
\end{center}
\end{figure}

\section*{References}
\begin{thereferences}

\item Aerts, C., et al., 1999, A\&A, {\bf 343}, 872

\item Aerts, C.; Marchenko, S. V.; Matthews, J. M.; Kuschnig, R.; Guenther, D. B.; Moffat, A. F. J.; Rucinski, S. M.; Sasselov, D.; Walker, G. A. H.; Weiss, W. W. 2006a, ApJ, {\bf 642}, 470

\item Aerts, C.; De Cat, P.; Kuschnig, R.; Matthews, J. M.; Guenther, D. B.; Moffat, A. F. J.; Rucinski, S. M.; Sasselov, D.; Walker, G. A. H.; Weiss, W. W. 2006b, ApJ., {\bf 642}, L165

\item Allende P. C., Lambert D. L. 1999, A\&A, {\bf 352}, 555

\item Barban, C.; Matthews, J. M.; de Ridder, J.; Baudin, F.; Kuschnig, R.; Mazumdar, A.; Samadi, R.; Guenther, D. B.; Moffat, A. F. J.; Rucinski, S. M.; and 3 coauthors 2007, A\&A, {\bf 468}, 1033

\item Bedding, T. R.; Kjeldsen, H.; Bouchy, F.; Bruntt, H.; Butler, R. P.; Buzasi, D. L.; Christensen-Dalsgaard, J.; Frandsen, S.; Lebrun, J.-C.; Martic, M.; Schou, J. 2005, A\&A, {\bf 432}, L43

\item Bruntt, H.; Kjeldsen, H.; Buzasi, D. L.; Bedding, T. R.

 2005, ApJ, {\bf 633}, 440

\item Cayrel de Strobel G., Soubiran C., Ralite N. 2001, A\&A, {\bf 373}, 159

\item Cunha, M.~S.\ 2001, MNRAS, {\bf 325}, 373

\item De Ridder J., Barban C., Carrier F., et al. 2006, A\&A, {\bf 448}, 689

\item Eggenberger P, Carrier F, Bouchy F and Blecha A 2004 {\it A\&A} {\bf
422} 247

\item Kallinger et al. 2007, A\&A, submitted September

\item Kjeldsen, H.; Bedding, T. R. 1995, A\&A, {\bf 293}, 87 

\item Kurtz, D.~W., et al.\ 2002, MNRAS, {\bf 330}, L57

\item Kurtz, D.~W., et al.\ 2005, MNRAS, {\bf 358}, 651

\item Leccia, S., Kjeldsen, H., Bonanno, A., Claudi, R. U., Ventura R and Patern\'o L
2007 {\it A\&A} {\bf 464} 1059

\item Lef\`evre, L.; Marchenko, S. V.; Moffat, A. F. J.; Chené, A. N.; Smith, S. R.; St-Louis, N.; Matthews, J. M.; Kuschnig, R.; Guenther, D. B.; Poteet, C. A.; Rucinski, S. M.; Sasselov, D.; Walker, G. A. H.; Weiss, W. W. 2005, ApJ, {\bf 634}, L109 

\item Marti\'c M, Lebrun J-C, Appourchaux T and Korzennik S G 2004 {\it A\&A}
{\bf 418} 295

\item
Matthews, J.M., Kusching, R., Guenther, D.B., Walker, G.A.H., Moffat, A.F.J., Rucinski, S.M., Sasselov, D., Weiss, W.W. 2004, Nature, {\bf 430}, 51

\item Reegen, P., Kallinger, T., Frast, D., Gruberbauer, M., Huber, D., Matthews, J. M., Punz, D., Schraml, S., Weiss, W. W., Kuschnig, R., J. Moffat, Walker, G. A. H., Guenther, D. B.,  Rucinski, S. M., Sasselov, D. 2006, MNRAS, 367, 141
\item Rowe, Jason F.; Matthews, Jaymie M.; Seager, Sara; Kuschnig, Rainer; Guenther, David B.; Moffat, Anthony F. J.; Rucinski, Slavek M.; Sasselov, Dimitar; Walker, Gordon A. H.; Weiss, Werner W. 2006, ApJ, {\bf 646}, 1241

\item R\'egulo, C., Roca Cort\'es, T.  2005, A\&A, {\bf 444}, L5

\item Richichi A., Percheron I., Khristoforova M. 2005, A\&A, {\bf 431}, 773

\item Robinson, F. J., Demarque, P., Guenther, D. B., Kim, Y.-C., and Chan,  
K. L. 2005, MNRAS, {\bf 362}, 1031

\item Saio, H.; Kuschnig, R.; Gautschy, A.; Cameron, C.; Walker, G. A. H.; Matthews, J. M.; Guenther, D. B.; Moffat, A. F. J.; Rucinski, S. M.; Sasselov, D.; Weiss, W. W. 2006, ApJ, {\bf 650}, 1111

\item Saio, H.; Cameron, C.; Kuschnig, R.; Walker, G. A. H.; Matthews, J. M.; Rowe, J. F.; Lee, U.; Huber, D.; Weiss, W. W.; Guenther, D. B.: Moffat, A.F.J; Rucinski, S.M.; Sasselov, D. 2007, ApJ, {\bf 654}, 544

\item Waelkens, C., 1991, A\&A, {\bf 246}, 453

\item Walker, G.A.H., Matthews, J.M., Kuschnig, R., Johnson, R.,
    Rucinski, S., Pazder, J., Burley, G., Walker, A., Skaret, K.,
    Zee, R., Grocott, S., Carroll, K., Sinclair, P., Sturgeon, D.,
    Harron, J. 2003, PASP, {\bf 115}, 1023

\item Walker, G. A. H. et al. 2005a, ApJ, {\bf 623}, L145

\item Walker, G. A. H.; Kuschnig, R.; Matthews, J. M.; Cameron, C.; Saio, H.; Lee, U.; Kambe, E.; Masuda, S.; Guenther, D. B.; Moffat, A. F. J.; Rucinski, S. M.; Sasselov, D.; Weiss, W. W. 2005b ApJ, {\bf 635}, L77

\end{thereferences}

\end{document}